\documentclass[10pt]{iopart}
\usepackage{graphicx}
\begin{document}

\title{The influence of magnetic order on the magnetoresistance anisotropy of Fe$_{1+\delta-x}$Cu$_{x}$Te}
\author{T. Helm$^{1,2}$\footnote{Present address: Max Planck Institute for Chemical Physics of Solids, Noethnitzer Str. 40, 01187 Dresden, Germany.}, P. N. Valdivia,$^3$, E. Bourret-Courchesne$^2$, J. G. Analytis$^{1,2}$, R. J. Birgeneau$^{1,2,3}$}

\address{Department of Physics, University of California, Berkeley, CA 94720, USA}
\address{Materials Sciences Division, Lawrence Berkeley National Laboratory, Berkeley, CA 94720, USA}
\address{Department of Materials Science and Engineering, University of California, Berkeley, CA 94720, USA}
\ead{toni.helm@cpfs.mpg.de}

%

\begin{abstract}
We performed resistance measurements on Fe$_{1+\delta-x}$Cu$_{x}$Te with $x_{EDX}\leq 0.06$ in the presence of in-plane applied magnetic fields, revealing a resistance anisotropy that can be induced at a temperature far below the structural and magnetic zero-field transition temperatures. The observed resistance anisotropy strongly depends on the field orientation with respect to the crystallographic axes, as well as on the field-cooling history. Our results imply a correlation between the observed features and the low-temperature magnetic order. Hysteresis in the angle-dependence indicates a strong pinning of the magnetic order within a temperature range that varies with the Cu content. The resistance anisotropy vanishes at different temperatures depending on whether an external magnetic field or a remnant field is present: the closing temperature is higher in the presence of an external field. For $x_{EDX} = 0.06$ the resistance anisotropy closes above the structural transition, at the same temperature at which the zero-field short-range magnetic order disappears and the sample becomes paramagnetic. Thus we suggest that under an external magnetic field the resistance anisotropy mirrors the magnetic order parameter. We discuss similarities to nematic order observed in other iron pnictide materials.
\end{abstract}



\submitto{\JPCM}

\maketitle

\ioptwocol

\section{Introduction}

Over the course of almost a decade a vast number of Fe containing compounds have been found to host a superconducting ground state. One proposed candidate as a potential host of superconductivity (SC) is the Chalcogenide FeTe~\cite{Mizugushi2009, Li2013, Subedi2017}. However, SC has only been observed in certain substitutions~\cite{Mizuguchi2009,Noji2010,Guo2015,bao, kata} and strained or doped thin film samples~\cite{Han2010,Nie2010,Ciechan2014} so far. One key issue has been to understand what the triggering parameters are and how magnetic and structural correlations are linked to the electronic ground state in those compounds.

A rich variety of magnetic and structural symmetries, but no SC, are known to occur as a function of the added interstitial iron content $\delta$ across the phase diagram of Fe$_{1+\delta}$Te compounds~\cite{fang, li, bao, rodri2, mizuguchi, koz}. At low values of $\delta\leq 0.09$ a monoclinic phase with commensurate bicollinear magnetic order occurs while at high values of $\delta$ the structure is orthorhombic with incommensurate helical magnetic order. Increasing $\delta$ was found to change the character of the low-temperature resistance significantly towards a more insulating behavior~\cite{fang, rodriguez, koz, resAnisUchida2, machida}. This effect was at first attributed to the changes in structure~\cite{fang} and later to changes in the nature of the scattering~\cite{resAnisUchida2, machida} and donor contribution~\cite{machida} caused by interstitial iron. Alternatively, it has been proposed that the presence of a spin gap contributes to the metallic state~\cite{rodriguez}.

A slight amount of copper substitution into Fe$_{1.1}$Te, making Fe$_{1.1-x}$Cu$_{x}$Te, has been shown to suppress (concomitantly) the magnetic and structural phase transition temperatures~\cite{jinshengsFTC, resAnisUchida2, hangdongWang, hundsMontgomery, neutron} while the magnetic order remains long-range and commensurate at $x = 0.04$~\cite{neutron}. At $x = 0.1$, only short-range incommensurate magnetism exists and no structural transition is observed by neutron diffraction~\cite{jinshengsFTC}. Both DC~\cite{jinshengsFTC, hangdongWang, neutron} and AC susceptibility~\cite{hangdongWang} measurements suggest a spin-glass ground state for copper substitutions at and above $x = 0.1$. Further studies showed that at $x = 0.06$, the material exhibits neither long-range commensurate nor short-range spin-glass order, but rather exhibits features intermediate between these two regimes.

Neutron scattering experiments revealed a strong coupling between the magnetic and structural order in the latter composition~\cite{neutron} (as in Fe$_{1+\delta}$Te compounds~\cite{li, bao, rodri2, rodriguez}). Copper substitution therefore suppresses the magnetism and leads to short-range magnetic correlations, similar to what is observed in superconducting selenium substituted samples~\cite{bao, kata}. In contrast, copper-free compounds with high interstitial iron content exhibit long-range helical order; thus addition of interstitial iron does not result in short-range magnetic order except close to the magnetic and structural phase boundaries as in Fe$_{1.12}$Te. Here short-range incommensurate magnetic order coexists and competes with long-range helical order of higher incommensurability~\cite{rodri2}. Structural phase coexistence in this region was also confirmed by x-ray diffraction measurements~\cite{koz}.    

With regard to transport, copper substitution increases the residual resistance ratio, which already by $x = 0.06$~\cite{hangdongWang, neutron} exceeds the highest values observed in samples of Fe$_{1+\delta}$Te, at $\delta = 0.14$~\cite{rodriguez, machida} and $0.22$~\cite{fang}. An enhancement of resistance anisotropy was predicted for the single domain state due to ferro-orbital order~\cite{turner}.

The comparison of basic transport measurements to results from neutron diffraction suggested changes in the electronic order with copper substitution~\cite{neutron}. The structural $(1,0,0)$ peak indicating an orbital ordering transition~\cite{fobes,neutron} was observed for $x=0.04$ but is absent for higher, $x= 0.06$ Cu content. Thus, if the structural distortion which gives rise to the $(1,0,0)$ peak is necessary for electronic order to occur, then such order is not possible as these degrees of freedom are absent in $x = 0.06$. Hence, an orbital ordering scenario would predict a significant decrease in resistance anisotropy on going from $x = 0.04$ to $x = 0.06$. However, ascribing a physical mechanism to the single-domain in-plane resistance anisotropy is complicated as such anisotropy can also arise from other intrinsic effects such as Hund's coupling between itinerant electron spins and local moments~\cite{hundsMontgomery} or from extrinsic effects such as anisotropic impurity scattering~\cite{resAnisUchida1, resAnisUchida2}. Both Hund's coupling and anisotropic impurity scattering have been claimed to be significant in giving rise to the resistance anisotropy in Fe$_{1+\delta}$Te: the original study suggested both effects are needed to describe the evolution of the resistance anisotropy with annealing~\cite{hundsMontgomery} while another group disputed this claiming that the resistance anisotropy is only due to impurity scattering~\cite{resAnisUchida1, resAnisUchida2}. In addition, an intrinsic mechanism based on the Fermi surface anisotropy has also been suggested to govern the resistance anisotropy in the electron-doped iron-pnictides~\cite{pnictideFermiSurface}. Orbital order implies, but is not necessary for the existence of such a Fermi surface anisotropy, which can in principle exist for any non-tetragonal structure.

Furthermore, neutron diffraction experiments on $x$ = 0.06 at $13.3\,$K revealed loss of magnetic intensity and a lowering of structural intensity upon warming. That suggested the attainment of a field-induced phase which could also be stabilized by cooling in field~\cite{neutron}.

Those findings motivated us to look for field-induced changes in the resistance. However, it is known that applying an in-plane magnetic field can induce partial detwinning as it was shown for BaFe$_{2}$As$_{2}$~\cite{chuField,nematic}. Thus we expected that field-induced detwinning might occur in the iron-chalcogenides which have a higher local moment per iron site~\cite{li, bao, rodri2}. Distinguishing field-induced phase transitions from field-induced detwinning using resistance anisotropy measurements is not trivial.

Herein, we present resistance measurements taken upon warming Fe$_{1+\delta-x}$Cu$_{x}$Te compounds with copper composition $0 \leq x \leq 0.06$ in the presence of in-plane magnetic fields. When fields of sufficient strength are oriented along either of the in-plane axes, we are able to observe two sequential features in resistance versus temperature for each of the compounds studied. We correlate these features with the changes observed by previous neutron scattering experiments~\cite{neutron}. The peculiar evolution of the angle-dependent magnetoresistance with temperature and Cu content suggests that the resistance anisotropy is closely linked to low-temperature magnetic and structural order. Our findings also suggest that the short-range magnetic order is linked to the field-induced electronic anisotropy. For $x_{EDX} = 0.06$ it appears below $T = 43\,$K which is far above the zero-field structural phase transition at $T_S = 28\,$K. Thus we uncover a number of field-induced changes in the magnetic, and/or electronic structure and we show that introducing a minor fraction of Cu to the compound has a huge influence on the phase diagram of Fe$_{1+\delta-x}$Cu$_{x}$Te.

\section{Experimental Methods}
\label{exp meth}
The growth technique and physical properties of  Fe$_{1+\delta-x}$Cu$_{x}$Te were described elsewhere~\cite{neutron}. Crystals were cleaved and cut into small pieces until a suitably thin and rectangular piece was obtained, with a typical size of $2\,$mm in length, $0.25\,$mm in width, and less than $0.05\,$mm in thickness. Low ohmic ($\leq 2 \Omega$) contacts were achieved by first sputtering gold onto a freshly cleaved surface and attaching $25\,\mu$m gold wires with EpoTek H20E silver epoxy. The sputtering pattern was four stripes across the width of the samples which ensures a uniform current profile. The samples were mounted with the $c$ axis perpendicular to the base of a Quantum Design parallel field rotator platform, so that the $ab$ plane is in the plane of rotation parallel to the magnetic field.  

All of the $R$ versus $T$ measurements shown were performed at a constant heating rate of $1.5\,$K/min, and additionally all of the cooling procedures which led into these warming measurements occurred at the same rate. All the measurements shown in the main text were taken after zero-field cooling (ZFC); except where otherwise noted. All $R$ versus $\phi$ measurements were performed in a stepping mode, that is, the data were recorded stepwise every 1$^{\circ}$ while the rotation was paused for 3 seconds; each angular position was set at a scan speed of $0.8$ degrees per second. In the present notation $\phi = 0^{\circ}$ corresponds to field along the $[1,0,0]$ direction.

When selecting samples for the measurements, we considered a large collection of crystals of cm dimensions. The copper composition for each of these samples was measured by Energy Dispersive X-ray Spectroscopy (EDX) thus we label the compounds using $x_{EDX}$. The measured interstitial iron content $\delta_{EDX}$ was $0.13$ for the doped samples except in $x_{EDX} = 0.05$ in which it was $0.15$. For the $x=0.06$ sample $\delta=0.13\pm0.01$ was confirmed by neutron powder diffraction (NPD) refinements. The copper free sample investigated in this work has $\delta=0.09\pm0.01$ confirmed by NPD and comparison to reference susceptibility and resistivity data~\cite{machida}. In the following we describe the samples using only $x_{EDX}$. $R$ vs. $T$ measurements were collected on these large crystal pieces. Following this, we cleaved the samples to create the rectangular specimen as previously described. In each composition, the $R$ vs. $T$ data from the rectangular specimen closely resembled that taken on the larger crystals confirming homogeneity of the samples found by EDX-measurements within mm to cm size volumes of the crystal boules as described previously~\cite{neutron}.

\section{Results}
\label{results}
\subsection{Resistance Versus Temperature Measurements}

\begin{figure}[tb]
\centerline{\includegraphics[width=0.9\linewidth]{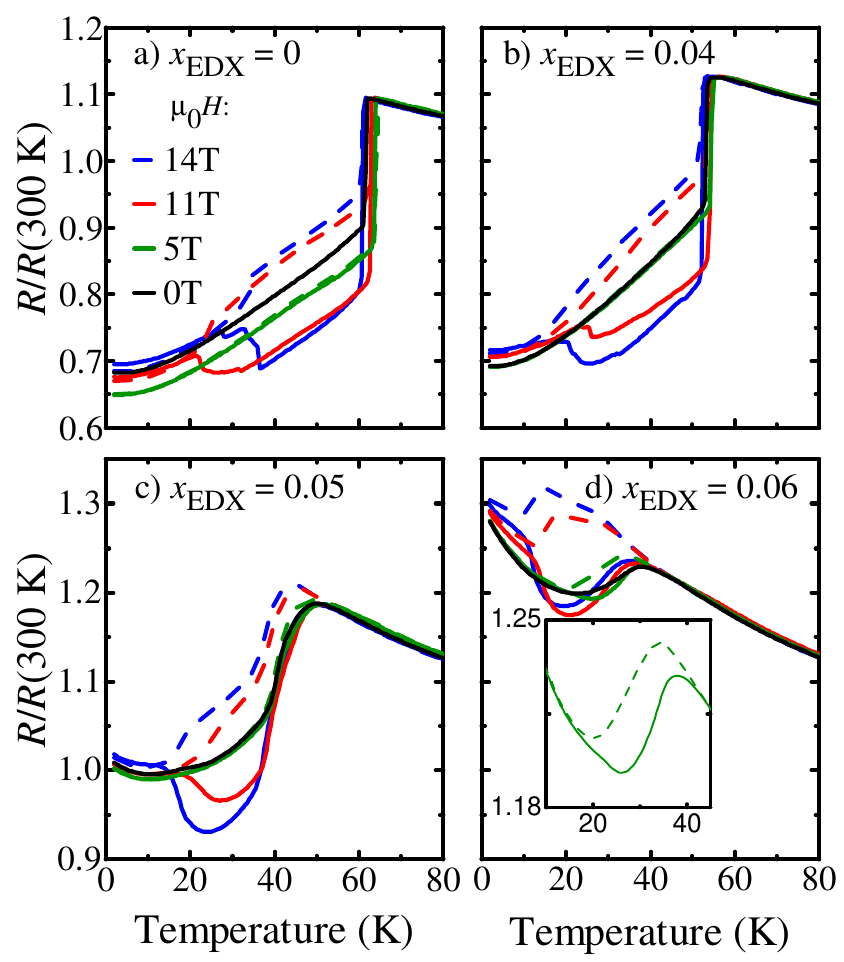}}
\caption{Resistance versus temperature for each composition in a variety of fields. Both $R_{H // [1,0,0]}$ (dash) and $R_{H // [0,1,0]}$ (solid) are shown at each field with the former having higher resistance than the latter. The inset in panel (d) shows a zoom-in of the $\mu_{0}H = 5\,$T data.}  
\label{F1}
\end{figure}

Figure~\ref{F1} shows the normalized resistance versus temperature data taken in a variety of fields for $x= 0$, 0.04, 0.05 and $0.06$. As noted above, we have fixed the reference frame for the crystal orientation such that the current, $I$, flows along the crystallographic $a$ axis, that is along $[1,0,0]$. At $T_N$, that is the temperature of the transition to the paramagnetic state, for low $x$ resistance experiences a sharp step-like increase right before it acquires a monotonic negative slope at high temperatures. This feature acquires a more gradual behavior for larger $x$. We observe a clear deviation between the resistance measured with the magnetic field, $\mu_0H$, parallel to $[1,0,0]$ and $[0,1,0]$: For a particular temperature range the resistance measured with $\mu_0 H // [0,1,0]$ (solid) is lower than that measured with the field along $\mu_0 H // [1,0,0]$ (dash). This in-plane resistance anisotropy evolves gradually upon warming until it suddenly closes in the vicinity of the magnetic transition temperature.

\begin{figure}[b]
	\centerline{\includegraphics[width=0.9\linewidth]{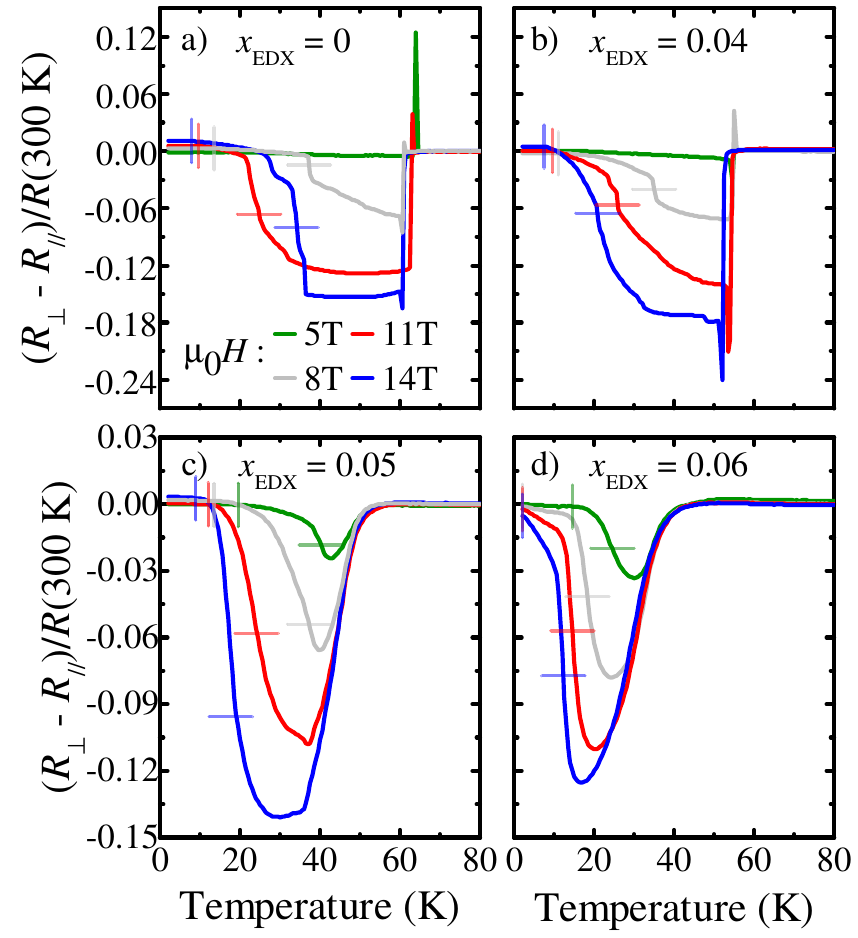}}
	\caption{The difference between two current configurations presented in Fig.~\ref{F1} i.e. $R_{H//[0,1,0]}$ - $R_{H//[1,0,0]}$: vertical lines denote $T_{1}$ while horizontal lines denote $T_{2}$}
	\label{F2}
\end{figure}

In Fig.~\ref{F2} we show an alternate representation, namely the difference between the two field-orientation curves, $R_{\perp}-R_{||}$. It develops a nonzero slope at $T_{1}$ (vertical marks) followed by a steep, step-like increase at higher temperatures. We label the temperature at which the derivative of the difference curve experiences a maximum using $T_{2}$ (horizontal marks in Fig~\ref{F2}). The inset of Fig.~\ref{F1} shows a zoom-in plot of the resistance curves for $x_{EDX} = 0.06$ in $\mu_{0}H = 5\,$T as this composition and field were examined in the neutron scattering experiment; a slight resistance anisotropy develops at $T_{1} = 14.5\,$K and lasts until $T = 43\,$K; these temperatures correspond closely to the magnetic and paramagnetic transition observed by neutron scattering experiments~\cite{neutron}, respectively, as we shall return to in the discussion, see Fig:~\ref{F9}.   

Considering any of the compounds shown in Fig.~\ref{F1} and ~\ref{F2}, it can be seen that in addition to a weak magnetoresistance induced by the field at base temperature (which was reversible upon removing the field if the samples had been cooled in zero-field), increasing the field will increase the resistance anisotropy when present. Furthermore, we find that $T_1$ and $T_2$ move to lower temperatures as the field is increased. However, we find that for $x_{EDX}=0$ the lowest onset temperatures are reached for $B\leq 11\,$T. The $14\,$T curve exhibits the highest anisotropy value but higher onset temperatures.

For $x_{EDX} = 0$ and $0.04$, the difference appears to settle at a minimum value at intermediate temperatures in high enough fields, suggesting that the parameters giving rise to the resistance anisotropy can reach an equilibrium in these field-induced states. We find that for $x_{EDX} = 0$ in $\mu_{0}H = 14\,$T there appear to be multiple sequential steps in the difference curve on raising the temperature above $T_{1}$. For increasing copper content those sharp features vanish almost completely.

\begin{figure}[tb]
	\centerline{\includegraphics[width=0.9\linewidth]{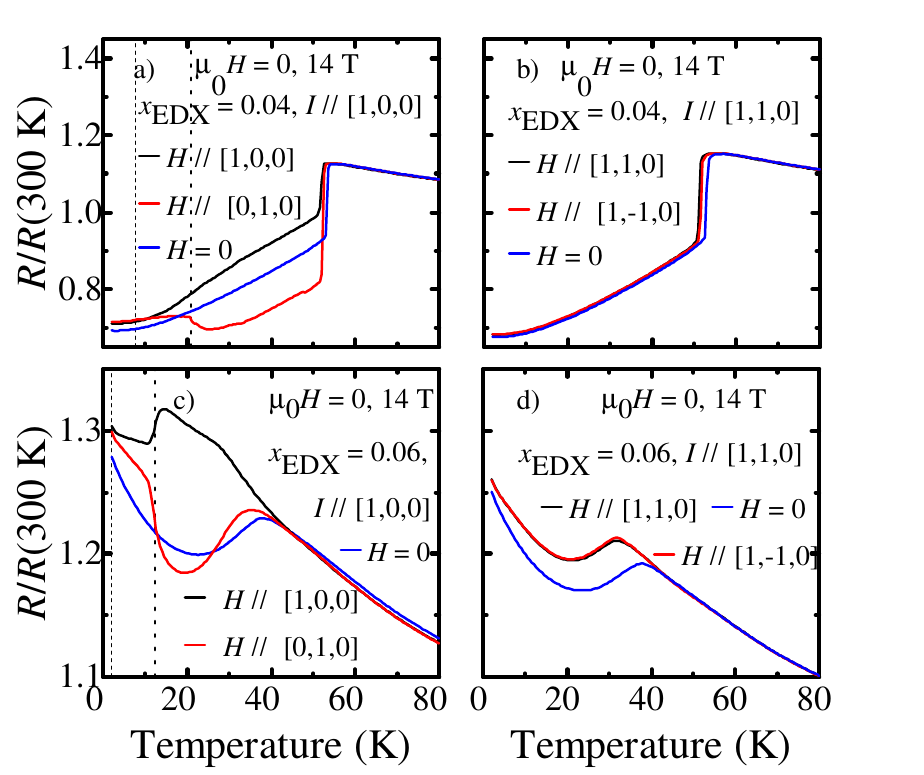}}
	\caption{$R$ vs $T$ measurements in $\mu_{0} H = 14\,$T comparing $I//[1,0,0]$ and $I//[1,1,0]$ for $x_{EDX} = 0.04$ and $0.06$. The dashed and dotted lines in panels (a) and (c) denote $T_{1}$ and $T_{2}$, respectively. }
	\label{F3}
\end{figure}

Furthermore, what is most important in Fig.~\ref{F2} is that the behavior of the resistance anisotropy as it closes is quite different from one compound to the next.  For $x_{EDX} = 0$ and $0.04$, the transition to paramagnetism is sharp and well-defined; in contrast, for $x_{EDX} = 0.05$ and 0.06 the resistance anisotropy decreases smoothly to zero over a wide range of temperatures. This is similar to the way that the resistance increases either quickly for $x_{EDX} = 0$ and 0.04 or gradually with temperature for $x_{EDX} = 0.05$ and 0.06 in zero field (see Fig.~\ref{F1}). In particular for $x_{EDX} = 0.05$ and for $0.06$, the difference curves at different values of field merge at temperatures at which the resistance anisotropy is still non-zero. We shall return to discuss the meaning of this finding in the discussion. In order to confirm that the anisotropic magnetoresistance relates to the field-crystal orientation, we investigated samples for which the current was applied along $[1,1,0]$. As shown in Fig.~\ref{F3}, these samples exhibit the same value of resistance whether the field is parallel or perpendicular to the current, unlike the case for samples with $I//[1,0,0]$.

\begin{figure}[b]
	\centerline{\includegraphics[width=0.85\columnwidth]{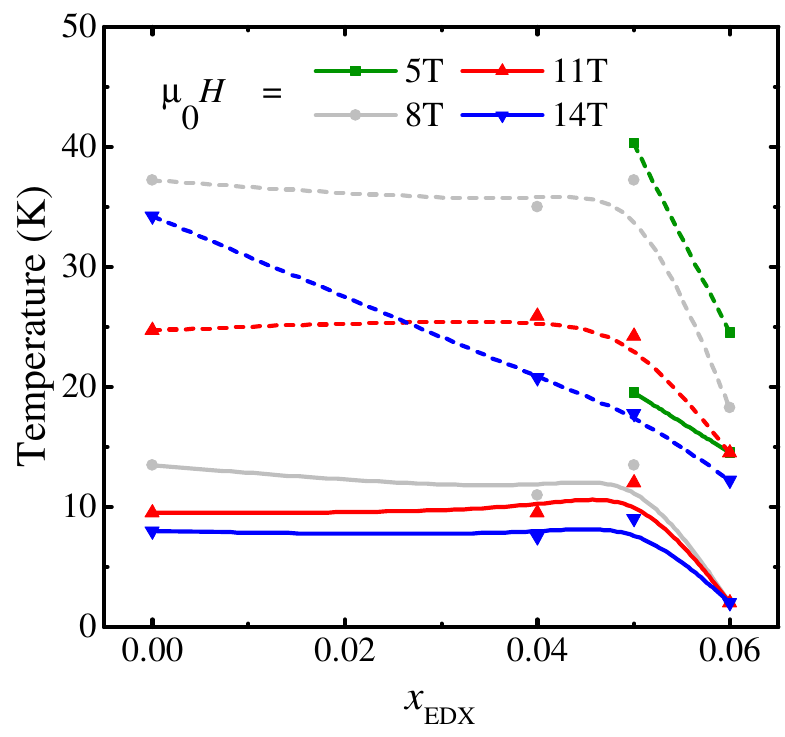}}
	\caption{Evolution of the resistance-anisotropy onset-temperature, $T_{1}$, (solid lines) and the steep step-like enhancement at $T_{2}$ with the Cu content $x$ (dashed lines). Lines are guides to the eye.}
	\label{F4}
\end{figure}

In Fig.~\ref{F4}, we show the resistance-anisotropy onset-temperature, $T_{1}$, and the step-like feature at $T_{2}$ (determined as the maximum of the derivative), versus $x_{EDX}$ for several field values: the lines are guides to the eye to help observe the general trends with field and composition. High magnetic fields are correlated with higher resistance anisotropies, as well as with a reduction in both $T_{1}$ and $T_{2}$. At $\mu_{0}H = 14\,$T, however, $T_{2}$ for $x_{EDX}$ = 0 seems to deviate from this general trend. However, this value is nearly equal to that obtained by a linear extrapolation from higher compositions at constant field. At this point we cannot identify the origin of the observed anomaly conclusively. It may suggest a different high-field regime exists.

\subsection{Effect of Field-Cooling (FC) Protocol on Resistance Anisotropy}

\begin{figure}[b]
	\centerline{\includegraphics[width=0.9\linewidth]{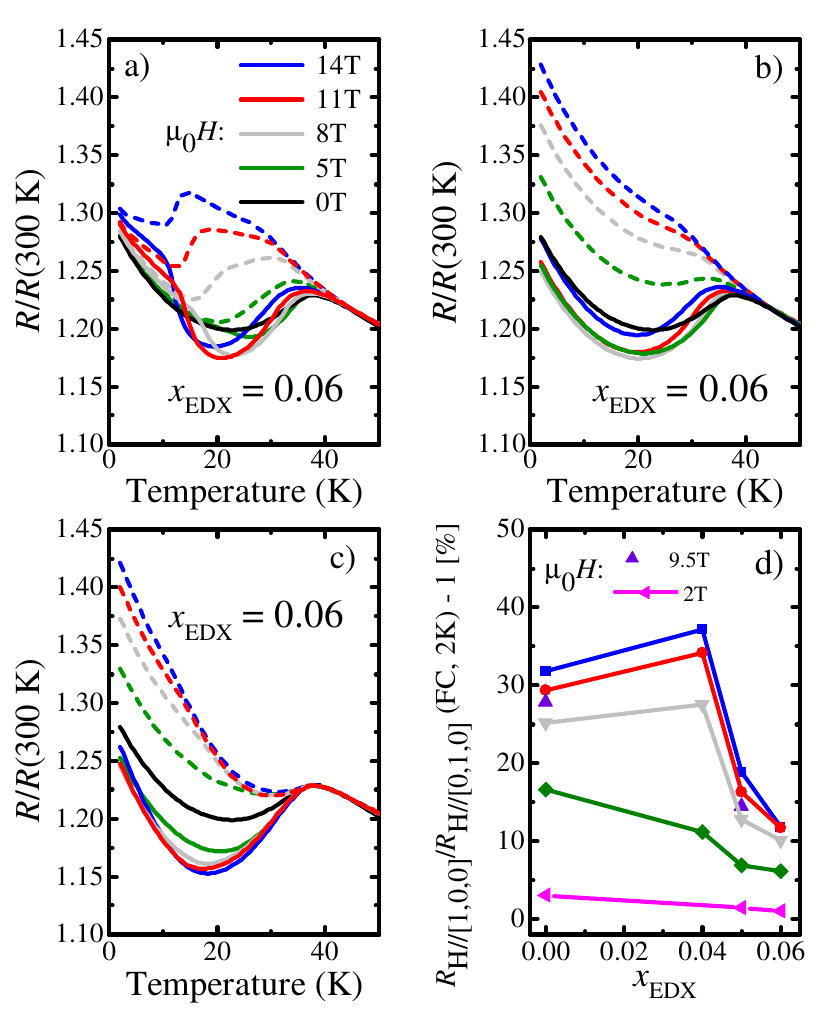}}
	\caption{Resistance $R$ versus the in-plane angle $\phi$ record upon warming after different cooling conditions (a) after ZFC (b) after FC, showing larger resistance anisotropy (c) after FC and then ramping the field to zero at $T = 2\,$K (remnant field measurement), showing the absence of anisotropy above the resistance maximum. Both $H//[1,0,0]$ (dash) and $[0,1,0]$ (solid) are shown at each field for (a)-(c). (d) The FC resistance ratio at $T = 2\,$K at various values of field for all compounds}
	\label{F5}
\end{figure}

In Fig.~\ref{F5}, we demonstrate on the example of $x_{EDX}=0.06$ that the magnitude of the field-induces resistance anisotropy depends strongly on the field-cooling (FC) protocol: Anisotropy observed for warming in field after ZFC (Fig.~\ref{F5}a) is even larger and extends to even lower temperatures when the cooling was performed in a non-zero field (see Fig.~\ref{F5}b and c). Fig.~\ref{F5}c indicates that in the absence of an applied field, there is no resistance anisotropy above the resistance maximum at $T=38\,$K in $x_{EDX} = 0.06$. In contrast, resistance anisotropy exists above the resistance maximum in the presence of the field, as shown in Fig.~\ref{F5}a and b for this compound. Field effects of the closing of the resistance anisotropy at the transition to paramagnetism are found for the other compounds as well, as can be seen in Fig.~\ref{F1}. Furthermore, the threshold field for obtaining resistance anisotropy by FC is lowered (in fact no minimum threshold field was found) -- resistance anisotropy could be observed in each compound at base temperature ($T = 2\,$K) after FC in  $\mu_{0}H = 2\,$T, as indicated in Fig.~\ref{F5}d.    

\subsection{Resistance vs field orientation measurements: shape and hysteresis effects}

\begin{figure}[tb]
	\centerline{\includegraphics[width=0.9\linewidth]{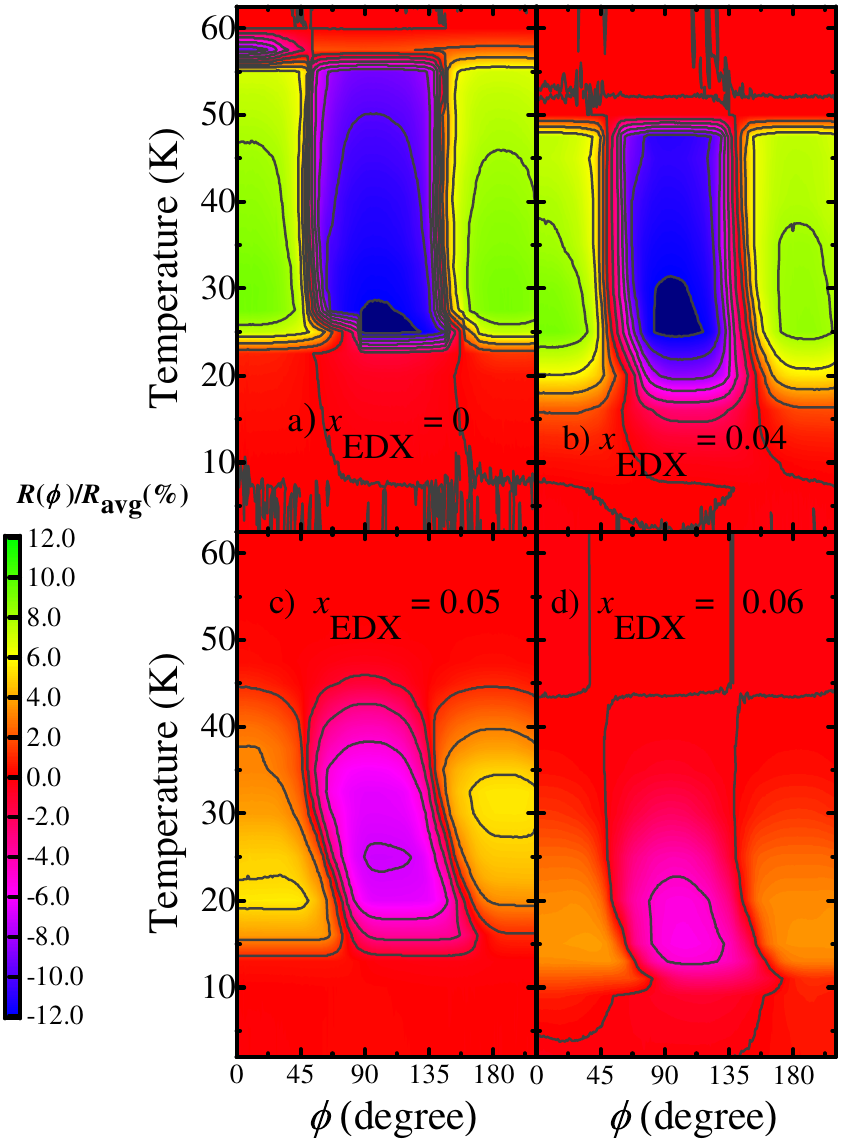}}
	\caption{Contour plots of the $R$ vs. $\phi$ data upon warming, for positive rotation (increasing $\phi$) in $\mu_{0}H = 14\,$T for $x_{EDX} = 0$ (a); 0.04 (b); 0.05 (c); 0.06 (d). Color represents the deviation from the average value at each temperature.}
	\label{F6}
\end{figure}
Fig.~\ref{F6} shows contour plots of the $R$ vs. $\phi$ measurements for each compound taken on rotating from low angle to high angle in a constant applied field of $\mu_{0}\textit{H} = 14\,$T. The measurements are performed by turning the field on after ZFC with the field parallel to [1,0,0], which we denote as $\phi = 0^\circ$, and then slowly scanning the angle to $\phi = 210^\circ$, then backwards to $\phi = 0^\circ$, and finally changing the temperature. The contour plots shown in Fig.~\ref{F6} consist only of unidirectional scans, i.e. increasing $\phi$. The extrema of the scans taken in each rotation direction are shifted from the high symmetry angles due to angular hysteresis effects which we shall discuss later. The color used in the contour plot represents the relative deviation of the resistance at each temperature from its average resistance over the full range of $\phi = [0-180]^\circ$.

While for samples with $x_{EDX} = 0$ and $0.04$, the angular positions of maximum and minimum resistance do not change significantly as the temperature is varied (i.e. the contour plots appear symmetric about their centers) we observe a monotonic shift of these positions for higher Cu content $x_{EDX} = 0.05$ and 0.06 samples.
Aside from this, much of the information encoded in these contour plots simply confirms the results of the $R$ vs. $T$ measurements above. As the Cu content increases, the resistance anisotropy decreases and closes near the paramagnetic transition temperature. While for $x_{EDX} = 0$ and 0.04  the transition from high to zero anisotropy is sharp and abrupt it becomes more and more gradual for larger $x$. We note that the weak negative magnetoresistance for $x_{EDX}=0$ at $57.5\,$K and small angles arises due to a slight temperature drift at the beginning of this particular rotation.

\begin{figure}[b]
	\centerline{\includegraphics[width=0.9\linewidth]{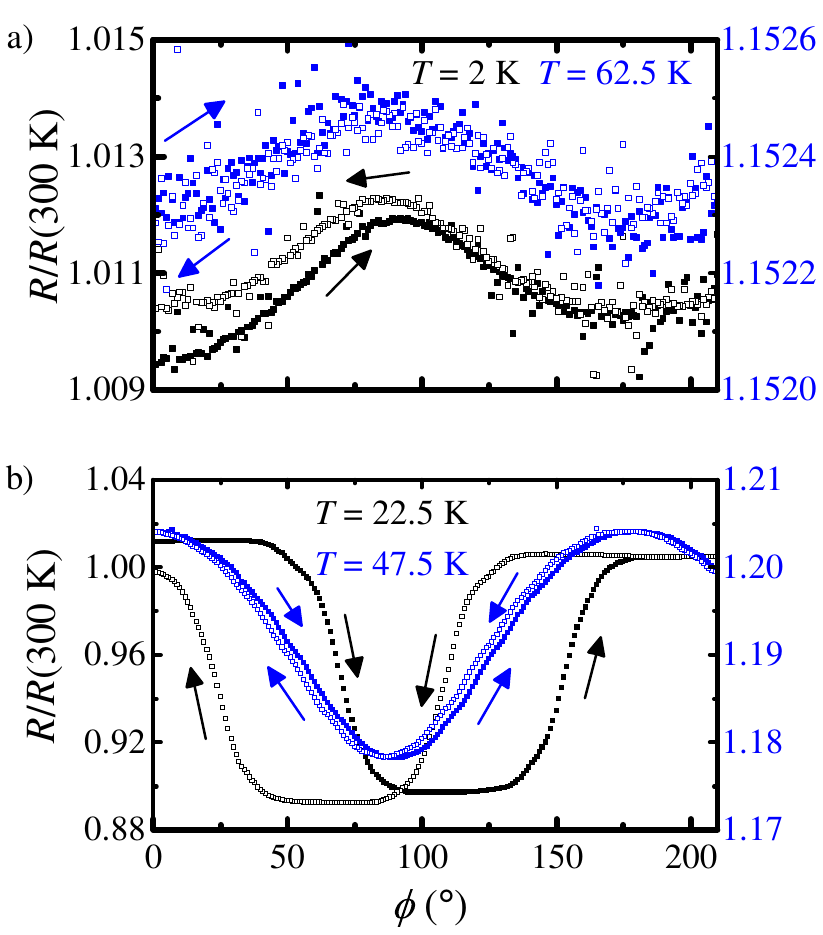}}
	\caption{\textit{R} vs. $\phi$ scans for $x_{EDX}$ = 0.05 in $\mu_{0}H$ = 14 T at \textit{T} = 2 K and 62.5 K (a); and at \textit{T} = 22.5 K and 47.5 K (b)}
	\label{F7}
\end{figure}
Fig.~\ref{F7} presents several rotation traces for $x_{EDX} = 0.05$ in $\mu_{0}\textit{H} = 14\,$T recorded at temperatures outside, that is below $T_1$ and above the structural/magnetic transition, (Fig.~\ref{F7}a) and within (Fig.~\ref{F7}b) the anisotropic region. In the former region the angle dependence of the resistance is very weak and follows $\cos(2\phi)$ with its minimum and maximum at field orientations close to $\mu_0 H$ parallel and perpendicular to the current, respectively. This weak variation can be attributed to ordinary magnetoresistance~\cite{citesDoring} due to the Lorentz force.

Fig.~\ref{F7}b shows $R$ vs. $\phi$ traces at two intermediate temperatures which are both above $T_{2}$ for $x_{EDX} = 0.05$. A sharp step-like transition between the high-resistance and low-resistance states occurs as a function of angle at lower temperatures (i.e. $T = 22.5\,$K) while at temperatures approaching the paramagnetic state (i.e. $T = 47.5\,$K) the magnitude of the $R$ vs. $\phi$ curve changes continuously with angle, very closely tracing a $\cos(2\phi)$ dependence. Also shown are data recorded on the way back, i.e. for the negative rotation. They demonstrate the large hysteresis observed in the measurements.
\begin{figure}[tb]
	\centerline{\includegraphics[width=0.9\linewidth]{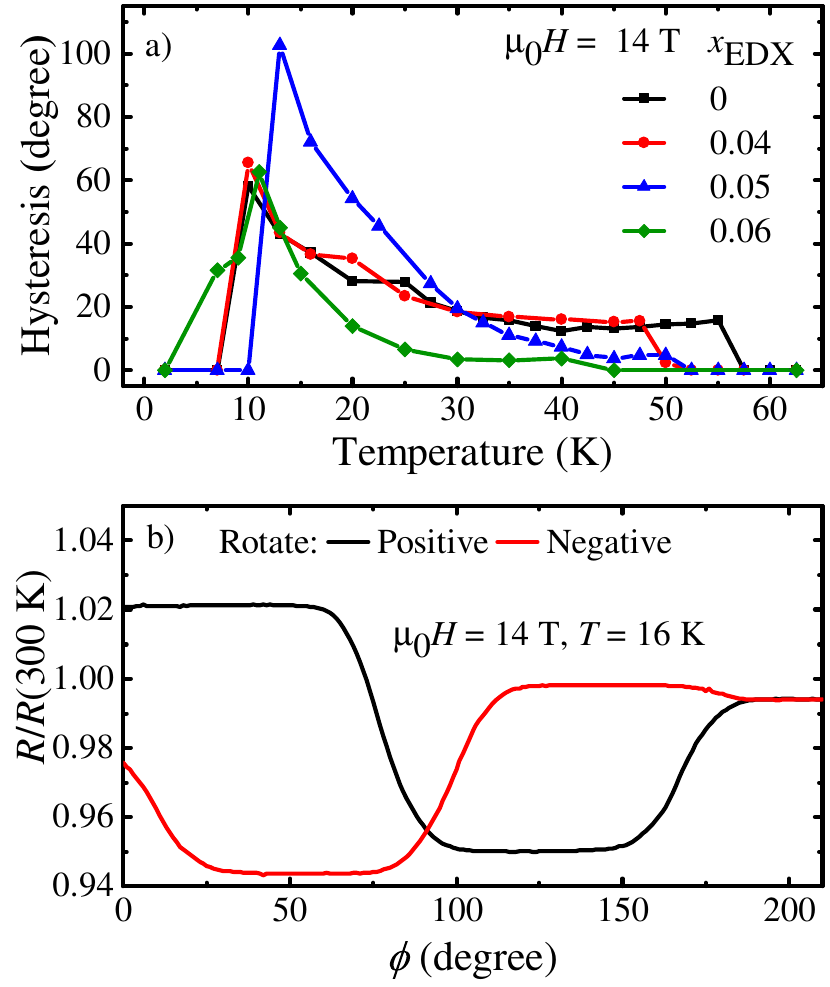}}
	\caption{(a) Hysteresis between positive and negative rotation of $R$ vs. $\phi$ measurements recorded in $\mu_{0}H =14\,$T. (b) Example  $14\,$T rotation curves for $x_{EDX} = 0.05$ at $T = 16\,$K.}
	\label{F8}
\end{figure}

The observed hysteresis can reach very large values, even above $90^\circ$ at low temperature for $x_{EDX} = 0.05$, implying a half phase shift of the positions of the maximum and minimum in resistance upon opposite rotations: an example scan with particularly high hysteresis of $72^\circ$ is shown in Fig.~\ref{F8}b. While for $x_{EDX} = 0$ and $0.04$ the hysteresis settles at a non-zero value at high temperature before abruptly jumping to zero, it gradually decays to very small values at high temperature for larger Cu contents $x_{EDX} = 0.05$ and $0.06$.

\section{Discussion and Conclusions}
\label{discu}  

The observed anisotropy in the temperature dependence of the resistance for Fe$_{1+\delta-x}$Cu$_x$Te, with $x=0$, $0.04$, $0.05$ and $0.06$ strongly depends on the field orientation with respect to the crystal lattice. We showed in Fig.~\ref{F3} that this effect is clearly decoupled from Lorentz force contributions to the magnetotransport and that it is only observable with a well defined current path along the in-plane crystal $a/b$ axes. Resistance anisotropy has been observed previously in Fe$_{1+\delta}$Te~\cite{hundsMontgomery, twoKinds, resAnisUchida1, resAnisUchida2} as well as selenium and copper substitutions thereof~\cite{resAnisUchida1, resAnisUchida2} detwinned by the application of uniaxial pressure. In the case of Fe$_{1.09}$Te in-situ x-ray measurements under uniaxial strain provided evidences for the resistance anisotropy being correlated to monoclinic domain population~\cite{twoKinds}. Field-induced detwinning was observed for BaFe$_{2}$As$_{2}$~\cite{chuField} and could be a possible mechanism responsible for our observations. The observed gradual onset of anisotropy at $T_1$ followed by step-like changes indicates strong pinning forces which subside by copper substitution.

\begin{figure}[b]
	\centerline{\includegraphics[width=0.85\linewidth]{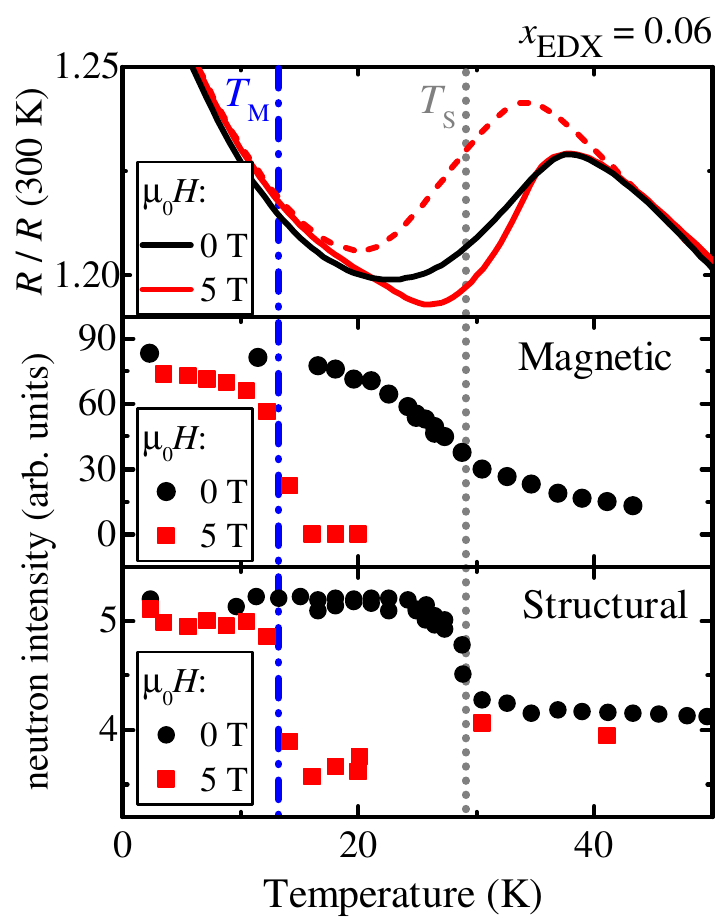}}
	\caption{$x_{EDX}=0.06$: Comparison of (a) the resistance with (b) the neutron scattering intensity of the magnetic $Q=(0.433,0,0.5)$ and (c) the structural $Q=(1,0,1)$ peak upon warming in zero field (black) and $\mu_{0}H=5\,$T after ZFC. Neutron data are taken from ref.~\cite{neutron}.}
	\label{F9}
\end{figure}
Previous neutron experiments, performed on $x_{EDX}=0.04$ samples, confirmed that the structural transition into the monoclinic phase occurs simultaneously with the onset of long-range, commensurate antiferromagnetism~\cite{neutron}. This is reflected in our data by the sharp closing of the resistance anisotropy. Similarly, the more gradual onset of structural as well as magnetic order, observed for $x_{EDX}=0.06$ by neutron scattering~\cite{neutron}, is reflected in the gradual closing of the resistance anisotropy. However, the low-temperature magnetic structure has not been determined conclusively. Comparison of our results on $x_{EDX} = 0.06$ in $\mu_{0}H = 5\,$T (see Fig.~\ref{F9}) to previous neutron experiments on the same composition in field provides evidence for field-induced structural and magnetic states~\cite{neutron}: $T_{1} (B=5\,{\mathrm T}) = 14.5\,$K corresponds to the magnetic transition temperature $T_{M}$ (dash-dotted line in Fig.~\ref{F4} and ~\ref{F9}) observed by neutron diffraction. $T_{2} (B=5\,$T$) = 24.5\,$K is also within the range of temperatures at which a second structural transition was determined to occur in the neutron experiment~\cite{neutron}. Apparently, the resistance anisotropy mimics the changes in the magnetic order and persists well above the field-free structural transition at $T_S$ (dotted line in Fig.~\ref{F9}). Future X-ray and neutron experiments could reveal the nature of those features and clarify the contribution of structural and magnetic rearrangements depending on the Cu concentration.

One hypothesis arising from the previous neutron observations was that the field-induced behavior in $x_{EDX} = 0.06$ might be due to random field effect of copper~\cite{neutron}. However, the observation of similar features in $x=0$ samples clearly indicates that substitution of copper is not a necessary ingredient for the field-induced behavior. Additionally, resistance anisotropy is discernible for temperatures up to $T = 43\,$K in $x_{EDX} = 0.06$. This suggests that some form of anisotropic order survives until this temperature, which corresponds precisely to the highest temperature at which short-range magnetic order was observed in zero field by neutrons~\cite{neutron}. Therefore it is likely that some form of magnetic order exists in the field-induced state of $x_{EDX} = 0.06$. By analogy, magnetic order induced by an external field may exist for $x\leq0.06$ as well.

Irreversible field-induced phase transitions have been previously observed by high-field susceptibility measurements in Fe$_{1.1}$Te albeit at much higher temperatures and applied fields~\cite{Knafo2013,Duc2014}. As a possible origin these authors suspected intrinsic magnetostructural changes~\cite{Paul2011} and detwinning. Similar behavior was observed by magnetoresistance measurements at high fields in $5\%$ and $10\%$ sulfur-substituted Fe$_{1+\delta}$Te~\cite{Tokunaga2012}. We note that our observed characteristic temperatures $T_1$ and $T_2$ occur at much lower fields compared to what was reported from these experiments. However, the in-plane field orientation is not clearly stated in the latter reference and it is not clear if current was aligned along a tetragonal crystal axis. Therefore, the latter results are somewhat ambiguous in the context of the present results which suggest that the field and current orientation as well as the field-cooling protocol is extremely important to observe these features (see the resistance data taken with $I//[1,1,0]$ in Fig.~\ref{F3} where a transition is not discernible). In addition, magnetoresistance was measured in Fe$_{1.05}$Te albeit with the field parallel to the $c$ axis~\cite{chenAnomaly} and thermoelectric in-plane transport on Fe$_{1.087}Te$~\cite{Pallecchi2009} and Fe$_{1.04}Te$~\cite{Matusiak2012}; no field induced anomaly was reported in these cases.

It has been argued that the antiferromagnetism in Fe$_{1+\delta}$Te is weak as inferred from heat capacity measurements; this is explained as resulting from competing ferromagnetic interactions which frustrate the antiferromagnetic order~\cite{zaliz}. Competition between different types of magnetic order has been implicated by neutron scattering experiments~\cite{rodri2, parshall, rodriguez} as well as experimental and theoretical pressure-dependence studies~\cite{Bendele2013,Monni2013}. Additionally, certain low-temperature orderings of in-plane structural bond-lengths have been suggested to play a role in the transport properties via the onset of electronic order~\cite{fobes}. It might be possible that a modification in the structural order to that observed by Fobes et. al in Fe$_{1.09}$Te~\cite{fobes} (hence resulting in ferro-orbital ordering) may take place when the atoms rearrange induced by external fields. Again future neutron and x-ray experiments could obtain further insights on that.

We consider that the high tunability of the resistance anisotropy with respect to changes in the field magnitude may either be due to a stronger depinning effect of increased fields or a variable volume fraction of a possible field-induced phase. If due to a variable volume fraction, then $T_{1}$ would correspond to a nucleation event which must occur homogeneously throughout the sample such that the correlation length of magnetic order (i.e. domain size) of the zero-field phase is effectively driven to zero when the field-induced phase first nucleates. This would be inevitable in order to account for the disappearance of magnetic intensity above $T_{M}$ observed by neutron diffraction as shown in Fig.~\ref{F9}.

We notice that the magnitude of the resistance anisotropy is quite similar for $x_{EDX} = 0$ and $0.04$, but is much smaller for $x_{EDX} = 0.05$ and $0.06$ (see Fig.~\ref{F6}). Hence, a dominant extrinsic impurity effect of copper substitution on the magnitude of the resistance anisotropy as was suggested by some authors~\cite{resAnisUchida1, resAnisUchida2} seems unlikely. Instead our findings suggest an intrinsic origin of the resistance anisotropy. However, we cannot distinguish between the different intrinsic mechanisms which might contribute to the resistance anisotropy i.e. Hund's coupling~\cite{hundsMontgomery} versus orbital ordering~\cite{turner} versus an anisotropic Fermi surface~\cite{pnictideFermiSurface} that does not involve orbital order or versus magnetostructural effects~\cite{Paul2011}.

Furthermore, unlike the lower $x_{EDX}$, the compounds with  $x_{EDX}=0.05$ and $0.06$ exhibit broad maxima in the temperature dependence of the zero field resistance. Additionally, a broad-in-temperature transitional region occurs between the resistance minimum and maximum in $x_{EDX} = 0.06$ in zero field. Also the zero-field resistance for $x_{EDX} = 0.05$ exhibits a broad increase between $T = 20\,$K and the resistance maximum. This stands in contrasts to the sharp, vertical increase in resistance at the paramagnetic transition for $x_{EDX} = 0$ and 0.04 (see especially Fig.~\ref{F1} and also ~\ref{F2} and~\ref{F6}). We interpret this as evidence that the primary (by which we mean the most prominent) transition in the structural order for $x_{EDX} = 0.05$ is suppressed relative to the onset of magnetic order in zero field, similar to the case of $x_{EDX}= 0.06$~\cite{neutron}. Thus it is likely that the smooth loss of resistance anisotropy with temperature (shown in Fig.~\ref{F2}) and the smooth loss of hysteresis with increasing temperature (shown in Fig.~\ref{F8}a) for the $x_{EDX} = 0.05$ and 0.06 compounds in field at temperatures approaching the paramagnetic state implies a continuous transition of the weak structural and/or magnetic orders in the field-induced state to zero. For comparison, we note that the hysteresis for the $x_{EDX} = 0$ and 0.04 compounds in field settles at a non-zero value for high temperature rather than decaying to zero (see Fig.~\ref{F8}a).

The sinusoidal shape of $R$ vs. $\phi$ scans at high temperatures in $x_{EDX} = 0.05$ and 0.06 indicates a weakened structural and/or magnetic order (see Fig.~\ref{F7}). Even at low temperatures for $x_{EDX} = 0.05$ and 0.06 the contour plots are asymmetric about their center unlike $x_{EDX} = 0$ and 0.04 (see Fig.~\ref{F6}). A sinusoidal angle dependence was also observed for BaFe$_{1.968}$Co$_{3.2}$As$_{2}$ at temperatures approaching the paramagnetic transition~\cite{chuField}. Notably, the antiferromagnetic transition is first-order at this cobalt composition~\cite{tricritical}. If the effects determining the shape of angular scans is similar for the two systems, then the change in shape will not be due to a continuous transition for the samples in the present study. In fact, the study on the pnictides~\cite{chuField} suggests that the maintenance of strong angular-switching behavior at temperatures approaching the paramagnetic transition in $x_{EDX} = 0$ and 0.04 is unique, which implicates either the double-stripe order or monoclinic symmetry in achieving this rigidity. We know that in zero field, there is a structural change between $x_{EDX} = 0.04$ and 0.06 in addition to the significant changes in magnetic ordering~\cite{neutron}. The zero-field structural and magnetic orders are quite different for these compounds. Thus, we expect that the field-dependent characteristics differ significantly too. The change from sharp to continuous switching observed in the $R$ vs. $\phi$ curve shape is similar to MnTe where a reduction in the antiferromagnetic domain size could explain those changes~\cite{jungwirthXavi}. Therefore, a reduced magnetic correlation length might allow for additional geometric degrees of freedom in the magnetic ordering pattern in the presence of applied fields. In the light of domains being influenced by the orientation of an external magnetic field our features at $T_1$ and $T_2$ might be regarded in terms of depinning temperatures that change depending on the field strengths and weakening of magnetic order due to changes in the Cu content. 

The resistance anisotropy in $x_{EDX} = 0.06$ closes at $T = 43\,$K, which is above the resistance maximum associated with the structural transition (see Fig.~\ref{F9}), but closes exactly at that maximum in the presence of the remnant field (see Fig.~\ref{F5}c). $T = 43\,$K corresponds to the highest temperature at which short-range magnetic order was observed by neutron diffraction for this compound~\cite{neutron}. In addition a clear cusp at $42\,$K in the magnetic susceptibility evidences a thermodynamic phase transition~\cite{neutron}. Consequently, it seems that the existence of short-range magnetic order is intimately connected with the occurrence of the resistance anisotropy. Phenomenologically, this looks similar to the electron-nematic state found in the iron-pnictides~\cite{nematic,fernandes2014}: a state with weak resistance anisotropy precedes a state with a higher resistance anisotropy that is correlated to a structural distortion. In our case the external magnetic field may play the role of a weak detwinning force that can induce resistance anisotropy. Since nematicity can be defined as broken rotational symmetry, the existence of short-range magnetic order may imply the possibility of resistance anisotropy. This provides a direct physical link between this chalcogenide and the iron-pnictides. However the analogy is not perfect, because as we have noted, the existence of short-range magnetic order implies the existence of a subtle structural distortion below $T = 43\,$K~\cite{neutron}. A similar behavior was observed under a high uniaxial pressure of $108\,$MPa with its origin in a softening of the lattice, possibly induced by orbital fluctuations. This effect was described by a divergent nematic susceptibility in Fe$_{1.09}$Te~\cite{twoKinds}. Experimental evidence for orbital fluctuations in Fe$_{1.13}$Te has been found by point contact spectroscopy measurements~\cite{arham} which led this group to posit the existence of resistance anisotropy above the structural transition in this compound.

\subsection{Conclusions}

In conclusion, we have shown that the resistive state of these iron-chalcogenide materials depends sensitively on the magnetic and structural orders, which can be modified by the application of a magnetic field within a certain temperature range. Already minor substitution of copper can have a strong effect on the softening of these orders. Our observations additionally suggest that subtle modifications of structural and/or magnetic order (when existent) might play a role in the properties of superconducting iron chalcogenides under sizable in-plane magnetic fields.

\ack
We are grateful to Meng Wang, Zhijun Xu, and Min Gyu Kim for experimental support. This work was supported by the Director, Ofﬁce of Science, Ofﬁce of Basic Energy Sciences, U.S. Department of Energy, under Contract No. DE-AC02-05CH11231, the Ofﬁce of Basic Energy Sciences U.S. DOE Grant No. DE-AC03-76SF008, and the Laboratory Directed Research and Development Program of Lawrence Berkeley National Laboratory under the US Department of Energy Contract No. DE-AC02-05CH11231. J.G.A. and T.H. also acknowledge the support of the Gordon and Betty Moore Foundation.

\end{document}